\pdfoutput=1 
\documentclass[a4paper,12pt]{article}
\usepackage{geometry}
\usepackage{lipsum} 
\usepackage[utf8]{inputenc}
\usepackage[T1]{fontenc}
\usepackage{bigstrut}
\usepackage{tikz}
\usepackage{tkz-euclide}
\usepackage{graphicx}
\usepackage{multirow,array}
\usepackage{mdframed}
\usepackage{mathtools}
\usepackage{multicol}
\usetikzlibrary{calc, matrix, arrows.meta}
\usepackage{zref-savepos}
\usepackage{amsmath,amssymb,amsthm, amsfonts}
\usepackage{tikz-inet} 
\usepackage{color}
\usepackage{relsize}
\usepackage{subcaption}
\usepackage{nicefrac}
\usepackage{apacite}
\usepackage{natbib}
\usepackage{mathrsfs}
\usepackage{fullpage}
\usepackage{comment}
\usepackage{float} 
\usepackage{setspace}
\theoremstyle{definition}

\theoremstyle{remark}

\newtheorem{definition}{Definition}
\usepackage{hyperref}
\usepackage[ruled,vlined]{algorithm2e}
\usepackage{authblk} 
\usepackage{titling}
\usepackage[symbol]{footmisc}
\usepackage{tabularx}

\title{An algorithm for two-player repeated games with imperfect public monitoring}
\author{Jasmina Karabegovic\thanks{Email: \texttt{jasmina.karabegovic@uni-graz.at}. I am very grateful to Christoph Kuzmics and Michael Greinecker for helpful discussions and constructive criticism. I gratefully acknowledge financial support from the Jubiläumsfond of the Österreichische Nationalbank (Project 18899).}}
\affil[]{Department of Economics, University of Graz}
\date{October 15, 2024} 
\newcommand{\keywords}[1]{\textbf{Keywords}: #1}

\begin{document}

\maketitle

\begin{abstract}
    This paper introduces an explicit algorithm for computing perfect public equilibrium (PPE) payoffs in repeated games with imperfect public monitoring, public randomization, and discounting. The method adapts the established framework by \citet*{abreu1990toward} into a practical tool that balances theoretical accuracy with computational efficiency. The algorithm simplifies the complex task of identifying PPE payoff sets for any given discount factor $\delta$. 
    A stand-alone implementation of the algorithm can be accessed on
    \href{https://github.com/jasmina-karabegovic/IRGames.git}{GitHub}.
\end{abstract}
 \medskip

\noindent  \keywords{Repeated games, imperfect public monitoring, computation}

\section{Introduction}
\label{intro:intro}
Games with imperfect monitoring play an important role in the theory of principal-agent problems (\cite{radner1985repeated}, \cite{rubinstein1983repeated}), collusion in oligopolies (\cite{green1984noncooperative} and \cite{athey2001optimal}), relational contracting (\cite{levin2003relational}) and governmental credibility (\cite{phelan2001sequential}). The Folk theorem for games with imperfect monitoring (\cite{0aa08f24-79b2-3e11-96e6-be22d23f2d5d}, see also \cite{abreu1986optimal, abreu1990toward}) suggests that for sufficiently patient players it is possible to sustain almost the entire set of (individually rational) feasible payoff profiles (from now on simply payoffs). Yet, the characterization of the equilibrium payoff set when the players are not as patient ($\delta < 1$) is not straightforward and there is, so far, no known way to solve this explicitly.

This paper provides an algorithm to compute the set of all perfect public equilibrium (PPE) payoffs for two-player repeated games with discounting, imperfect public monitoring, pure public strategies, and public randomization. 
The algorithm computes the set of normalized repeated game equilibrium payoffs as a function of the, assumed common, discount factor $\delta$ (ranging from $0$ to $1$). This is particularly important when considering impatient players, in which case the equilibrium payoff set is typically a strict subset of the feasible (and individually rational) set of payoffs.

The starting point in the literature is the conceptual algorithm by \cite{abreu1986optimal, abreu1990toward}, henceforth APS. 
They develop set-valued techniques for solving repeated games with imperfect monitoring. The APS algorithm works iteratively, starting with the stage game's feasible (and individually rational) payoff set. They demonstrate that the set of public perfect equilibrium payoffs could be characterized as the largest fixed point of a monotone set operator, resembling the Bellman operator in dynamic programming. This operator takes a set of payoffs that are possible at the next time step and delivers the payoffs that are possible at the current time step taking into account the incentive constraints imposed by the stage game. \cite{cronshaw1994strongly} apply the APS method to identify the set of strongly symmetric subgame perfect equilibrium payoff profiles in infinitely repeated games with discounting and perfect monitoring.
Identifying or even approximating the full equilibrium payoff sets for general repeated games for any given discount factor $\delta$ presents a significant challenge, as demonstrated, for instance, by Mailath et al.(\citeyear{mailath2002maximum}), who show that the problem becomes intractable even in the classic repeated prisoners’ dilemma with perfect monitoring.
While the APS techniques established the basis for calculating equilibrium payoff sets, the actual implementation of this method is not straightforward, as it requires operations on complex sets of payoffs that arise in later iterations of the algorithm. 

\cite{judd2003computing} overcome these problems by developing a numerical approach for computing subgame perfect equilibrium in infinitely repeated games with \emph{perfect} monitoring, using public randomization, by approximating payoff sets with convex polytopes. 
This allowed them to provide inner and outer approximations with computable error bounds on the equilibrium payoff set.
\cite{abreu2014algorithm} introduce an algorithm for approximating subgame-perfect equilibrium payoffs in two-player repeated games with, again, \emph{perfect} monitoring, by severely restricting the number of extreme points of the payoff set at any iteration of the algorithm. 
The insights provided by their algorithm allow them to solve analytically for the equilibrium payoff set. 
 
The algorithms mentioned so far were developed only for the case of perfect monitoring. \cite{goldlucke2012infinitely} provide an explicit algorithm for infinitely repeated games with imperfect public monitoring. However, they allow for observable monetary transfers. For such games, they are able to calculate the set of all public perfect equilibrium (PPE) payoffs.

This paper provides an algorithm to compute the set of all public perfect equilibrium (PPE) payoffs for two-player repeated games with discounting, imperfect public monitoring, and public randomization without monetary transfers.
Specifically, the proposed algorithm adapts the conceptual framework by APS into an explicit algorithm capable of computing (at least an upper bound for) the PPE payoff sets. The (computationally efficient) implementation is tailored to two-player games with up to three actions per player and up to four signals. 
The APS algorithm operates on complex convex sets with a rapidly increasing (in iterations) number of extreme points. The challenge encountered here is not just to find a way to approach these sets, but also to use a computationally efficient way to do so.

The paper is structured as follows: \hyperref[sec:repeated_games]{Sections 2} and \hyperref[sec:aps]{3} provide a comprehensive background on repeated games and the APS algorithm. \hyperref[sec:key_iteration_step]{Section 4} describes the key iteration step of the algorithm, and \hyperref[sec:cooperation_imperfect_monitoring]{Section 5} presents computed examples to illustrate the concepts discussed. 
\hyperref[sec:discussion]{Section 6} summarizes the main findings and the contribution of the paper to the literature.
Additionally, a stand-alone implementation of the algorithm can be accessed on
\href{https://github.com/jasmina-karabegovic/IRGames.git}{GitHub}.

\section{Repeated Games with Imperfect Public Monitoring}
\label{sec:repeated_games}
Two players, labeled 1 and 2, repeatedly play a finite stage game at all times $t \in \{0,1,2,...\}$. At each stage, each player $i$, simultaneously chooses an action from a finite action space $A_i$. The set of action profiles is given by $A = A_1 \times A_2$. At each stage, players receive a payoff given by $u_i:A \to \mathbb{R}$. For any $a \in A$, let $u(a)=(u_1(a),u_2(a)) \in \mathbb{R}^2$ denote the induced stage game payoff (profile). The set of stage game payoffs generated by the pure action profiles is
\[\mathscr{F} \equiv \{v \in \mathbb{R}^2: \exists a \in A \mbox{ s.t. } v = u(a)\}. \]
The set of feasible payoffs 
    \[W^* \equiv \text{co}\mathscr{F},\]
    is the convex hull of points in $\mathscr{F}$, which includes not only the payoffs directly achievable through pure strategies but also any payoffs that can be constructed as convex combinations of these points, representing mixed or randomized strategies.

\begin{definition}
\label{def:minmax}
    A player $i$'s minmax value is then given by
\[\underline{v}_i \equiv \text{min}_{a_{-i}\in A_{-i}} \text{max}_{a_i \in A_i} u_i(a_i, a_{-i}),\]
where, for $i \in \{1,2\}$, $-i$ denotes the other player, that is player $3-i$ (or $-i=1$ if $i=2$ and $-i=2$ if $i=1$). It is the minimum guaranteed payoff that player $i$ can secure regardless of the other player's actions.
\end{definition} 

\begin{definition}
\label{def:feasible_payoff}
The payoff vector $v=(v_1,v_2)$ is individually rational if $v_i \geq \underline{v}_i$ for each player $i$; it is strictly individually rational if the inequality is strict, i.e., $v_i > \underline{v}_i$, for each player $i$. 
    The set of individually rational and feasible payoffs is, then, given by
\[W^0 = \{v \in W^* \mid v_i \geq \underline{v_i} \text{ for all } i\}.\]
\end{definition}

\medskip

Repeated game payoffs are given by the normalized net present value of the flow of stage game payoffs. 
Players discount future payoffs by a common discount factor $\delta$. For any sequence of action profiles $a_0,a_1,a_2,... \in A$, this payoff is given by 
\[
(1-\delta) \sum_{t=0}^{\infty} \delta^{t} u(a_t).
\] \medskip

After each round $t$, a publicly observed signal $y$ is realized with the probability $\rho (y \mid a)$ from a finite signal space $Y$. 
No information beyond the signal $y$ (and their own actions) is available to the players. 
The set of public histories is given as 
\[\mathscr{H}\equiv \bigcup _{t=0} ^\infty Y^t.\]

\begin{definition}
    The set of public pure strategies for player $i$ is given as
\[
\sigma_i: \mathscr{H} \rightarrow A_i.
\]
\end{definition}

If players condition actions only on public histories, the continuation game is strategically the same as the original game, preserving the recursive structure of the equilibrium payoff set. 
Also, no player has an incentive to deviate to a non-public strategy when their opponent uses a public strategy (\cite{mailath2006repeated}, lemma 7.1.1). 

The approach in this paper allows for public randomization (at each stage). At the beginning of each period, the realization of some random variable is commonly observed by all players. Players condition their actions on the realized outcome. As is very common in the repeated games literature, see e.g., the textbook by \cite{mailath2006repeated}, it is assumed that this publicly observable random variable is distributed in such a way that players can (by conditioning on its outcome) generate any arbitrary joint distributions of action pairs. The assumption, therefore, convexifies the set of equilibrium payoffs, potentially allowing for simpler strategies to support equilibrium payoffs rather than a complicated sequence of pure actions (\cite{mailath2006repeated}, page 76; \cite{aumann1974subjectivity}). 

The analysis centers on Perfect Public Equilibria (PPE) in pure strategies. 
\begin{definition}
    A PPE is a profile of public strategies $\sigma=(\sigma_1,\sigma_2)$ that, for any public history $h^t$, specifies a Nash equilibrium of the repeated game.
\end{definition}
The aim is to calculate $\mathscr{E}(\delta) \subseteq W^0,$ the set of PPE repeated game payoffs given discount factor $\delta$.

The analysis is limited to only public pure strategies in the repeated game, focusing on PPE. 
According to lemma 7.1.2. in \cite{mailath2006repeated}, every pure strategy is equivalent to a public pure strategy.
While restricting the strategy space to public histories is a non-trivial assumption, it is essential to note that introducing private history dependence can result in equilibria with payoffs outside of the PPE payoff set (Mailath et al. (\citeyear{mailath2002private}), \cite{kandori2006efficiency}).
However, if the strategy space is limited to pure strategies, private histories become irrelevant, as they do not introduce new equilibria to the set (\cite{mailath2006repeated}, page 329).
This modeling choice is made for clarity and to preserve the recursive structure of the equilibrium payoff set.  
It makes it more tractable for the analysis, which would be otherwise more complex. 
The models under consideration do not necessitate conditioning on private histories.

\section{Abreu, Pearce, Stacchetti}
\label{sec:aps}
APS provide a method for characterizing the set of PPE payoffs in these games. This method is based on a set-valued dynamic-programming inspired approach that maps any set of possible continuation payoff profiles to the set of current, incentive compatible, payoff profiles. 

A payoff profile $v \in W^*$ is \emph{enforceable} \footnote{Terminology is used as in \cite{mailath2006repeated}.} given (continuation) payoff set $W \subseteq W^*$ if there exists a continuation payoff mapping $\gamma: Y \rightarrow W$ and an action profile $a \in A$ such that
\[v_i=(1-\delta)u_i(a) + \delta \sum_{y \in Y} \gamma_i(y) \rho(y \mid a),\]
for every player $i$, and

\[v_i \geq(1-\delta)u_i(a_i',a_{-i})+\delta \sum_{y \in Y} \gamma_i(y) \rho(y \mid (a_i',a_{-i})) \hspace{3em }\mbox{(IC)},\] 
for every player $i$ and every action $a_i$.

In words, a payoff profile $v \in W^*$ is \emph{enforceable} given (continuation) payoff set $W$, if there is a suitably chosen feasible continuation promise (given by $\gamma$)  in $W$ and a suitably chosen action profile $a \in A$ such that, first, payoff profile $v$ is achieved given $a$ and $\gamma$, and second, action profile $a$ is incentive compatible, that is no other action would give any player a higher current expected payoff given $\gamma$.

For any $W \subseteq W^*$, denote by $\mathcal{B}(W)$ the set of all convex combinations of enforceable payoff profiles given $W$. One of the main theorems of APS is that the equilibrium payoff set $\mathscr{E}(\delta)$ is then given as the largest fixed point of the mapping $\mathcal{B}$. Equivalently, it is given as the intersection of all generated sets
\[\mathcal{E}({\delta)} = \bigcap _{k = 1} ^ {\infty} \mathcal{B}^{k} (W^0),\] 
where, for any set $W \subseteq W^*$, $\mathcal{B}^{1}(W)=\mathcal{B}(W)$ and for $k \ge 2$, $\mathcal{B}^{k}(W)=\mathcal{B}\left( \mathcal{B}^{(k-1)}(W)\right)$.

\section{The Algorithm}
\label{sec:key_iteration_step}

The key step in the algorithm is the application of the set operator $\mathcal{B}$ to any convex set $W \subset W^*$. 
How this is programmed is described in some detail below (\hyperref[alg:the_alg]{see Table 1 below)}. Once we have this step, the algorithm starts with the set of feasible and individually rational payoffs and then iteratively applies the set operator $\mathcal{B}$, the key step of the algorithm, until either there is convergence, i.e., until we find a payoff set $W$ such that $\mathcal{B}(W) = W$, or until we find a set $W$ such that $\mathcal{B}(W)$ is as close to $W$ as we desire (see \hyperref[alg:the_alg1]{Table 2}). In the latter case, the algorithm produces only an upper bound on the equilibrium payoff set. 

Now to the key step of the algorithm. Let $W$ be some closed, bounded, and convex payoff set $W$. 
For each $a \in A$, the algorithm computes the expected average payoffs that can be obtained when $a$ is played and when continuation payoffs after each possible public signal (generated by $a$) are taken from $W$ and are such that the linear inequality incentive constraints, denoted (IC) above, are satisfied. Denote this set of obtainable payoff profiles using action profile $a$ by $P(a)$. The set $\mathcal{B}(W)$ can then be obtained as the convex hull of the union of all sets $P(a)$:

\[\mathcal{B}(W) = \mbox{co} \bigcup _{a \in A} P(a). \]

Given the assumption of allowing public randomization, all payoff sets are fundamentally convex polytopes. As the key trick to make the iteration steps as fast as possible, we use the fact that convex polytopes can be represented in two different ways: either as an intersection of a finite number of half-spaces or as the convex hull of a finite number of extreme points. 
The double-description method by \cite{MotzkinRaiffaThompsonThrall+1953+51+74} allows us to smoothly switch between these two representations.

The algorithm takes any input set $W$ in its minimal extreme point representation and then computes its half-space representation. It computes the continuation payoff sets for all $a \in A$ such that the IC is satisfied. It then computes average discounted payoff sets $P(a)$. 
The algorithm finally produces the extreme points of $\mathcal{B}(W)$, after finding the convex hull of the union of all $P(a)$.    

Going back and forth between the two representations, allows us to eliminate redundant inequalities, simplify the computations, and increase the algorithms' efficiency. 

The algorithm computes the area of all iterated payoff sets. The algorithm stops when the area difference between iterations $k$ and $k+1$ falls below the specified error bound $\epsilon$. Up to this point, the iterated payoff sets are calculated exactly. Only the fact that we need to stop the algorithm at some finite point implies that the payoff set that is the ultimate outcome of the algorithm provides an upper bound of the equilibrium payoff set and not the equilibrium payoff set itself. 

An inherent challenge in APS is the exponential growth in the number of extreme points, which impedes computational efficiency. To mitigate this, the algorithm allows for the option to integrate a key technical feature: the application of the Ramer-Douglas-Peucker (RDP) (\cite{ramer1972iterative}, \cite{douglas1973algorithms}) simplification algorithm. 
The RDP algorithm simplifies the piecewise linear boundaries of the polytope, effectively maintaining the geometric shape of the payoff set, while simplifying the lines that define it. 
Additionally, it offers the flexibility to set a specific threshold $\theta$ that meets the precision requirements. The threshold parameter $\theta$ determines the degree of simplification, ensuring that the essential characteristics of the payoff set are retained while unnecessary computational complexity is reduced. 
While the RDP feature enhances computational efficiency, there is a potential loss of accuracy, which depends on $\theta$, such that $\theta \ge 0$. For $\theta = 0$, RDP is off, no simplifications take place and the computations are exact. If $\theta$ increases, the approximations of the payoff sets are less accurate.

\begin{table}
    \begin{algorithm}[H]
    \DontPrintSemicolon
    \SetAlgoNlRelativeSize{-1}
    \SetNlSty{textbf}{}{}
    \SetAlgoNlRelativeSize{-2}
    \SetAlCapNameFnt{\small}
    \SetAlCapFnt{\small}
    \small 
    \textbf{Input:} V-representation of the set $Z = \{z_1, ..., z_m\}$ such that  $W = \text{co}(Z)$.

    \textbf{Step 1:} Compute the H-representation of $W$. The H-representation is in matrix form.
    
    \textbf{Step 2:} For each action profile $a$, add the corresponding incentive constraints to the H-representation of the payoff set by concatenating (stacking) the two matrices (one matrix for the H-representation and one for the incentive constraints).
    
    \textbf{Step 3:} For each action profile $a$ determine the set $P(a)$ of all vectors $v_i(a)$ such that for each player $i$, the following holds:
    \begin{align*}
     \quad & v_i = (1-\delta)u_i(a) + \delta \sum_{y} p(y \mid a)\gamma(y) \\
    \text{subject to:} \quad & v_i \geq (1-\delta) u_i(a') + \delta \sum_{y} p(y \mid a')\gamma(y) \quad \forall a' \in A, \\
    & \gamma(y) \in W.
    \end{align*}
    
    \textbf{Step 4:} Take the union of all the sets $P(a)$ calculated for every action profile to form a new set $W^{'}$.
    
    \textbf{Step 5:} Compute $\mathcal{B}(W)$ as the convex hull of $W^{'}$, and then obtain the V-representation $Z$ of this set, which is the output of this step. 
    
    \caption{The key iteration step}
    \label{alg:the_alg}
    \end{algorithm}
\caption{The key iteration step}
\end{table}

\begin{table}
    \begin{algorithm}[H]
    \DontPrintSemicolon
    \SetAlgoNlRelativeSize{-1}
    \SetNlSty{textbf}{}{}
    \SetAlgoNlRelativeSize{-2}
    \SetAlCapNameFnt{\small}
    \SetAlCapFnt{\small}
    \small 
    \textbf{Input:} V-representation of the individually rational feasible set $Z = \{z_1, ..., z_m\}$, where the individually rational feasible payoff set is $W = \text{co}(Z)$.

    \textbf{Initialization:} Set $k = 0$.
    
    \textbf{Iteration:}
    \begin{enumerate}
        \item Apply the key iteration step to $Z^k$ to obtain $Z^{k+1}$.
        \item Increment $k$ by 1.
    \end{enumerate}
    
    \textbf{Repeat} the iteration until the set induced from $Z^{k}$ converges (or the algorithm stops), ensuring that convergence is also determined by the specified threshold $\epsilon$ on the area difference of payoff sets $W^k$ and $W^{k+1}$ (the convex hull of $Z^k$ and $Z^{k+1}$, respectively).
    
    \caption{The algorithm}
    \label{alg:the_alg1}
    \end{algorithm}
\caption{The algorithm iteratively applies the key iteration step from Table 1.}
\end{table}

\newpage
\section{Implementation, Examples, and Evaluation}
\label{sec:cooperation_imperfect_monitoring}

In this section we illustrate the algorithm with a simple example inspired by the prisoners' dilemma, followed by a Cournot duopoly example, as presented in \cite{judd2003computing} and \cite{abreu2014algorithm}.

\paragraph{Implementation} We first need to define the parameters of the game: the payoff matrix, the discount factor, and the signal structure. The algorithm then identifies the stage game Nash equilibria, calculates the minmax payoffs, and computes the incentive constraints.

\subsection{A prisoners' dilemma example} Consider an infinitely repeated prisoners' dilemma with imperfect public monitoring, adapted from \cite{mailath2006repeated}, with the following payoff matrix:

\begin{table}[h]
    \centering
    
    \begin{center}
    \begin{tabular}{c|ccc}
     & C & D \\
        \hline
        C & $2, 2$ & $-1, 3$\\
        
        D & $3, -1$ & $0,0$ \\
                
    \end{tabular}
    \end{center}    
    
    \caption{The prisoners' dilemma}
    \label{fig:payoff_matrix}
\end{table}

At the end of each round, players observe one of two possible signals - the good signal $\Bar{y},$ or the bad signal $\underline{y}$. The signal structure is as follows: \bigskip

\[
    \rho(\Bar{y} \mid a) =
    \begin{cases}
    \frac{2}{3}, & \mbox{if}\ a=CC
    \\
    \frac{1}{2}, & \mbox{if}\ a=CD\  \mbox{or}\  DC
    \\
    \frac{1}{4} & \mbox{if}\ a=DD.
    \\
    \end{cases}
\] \bigskip

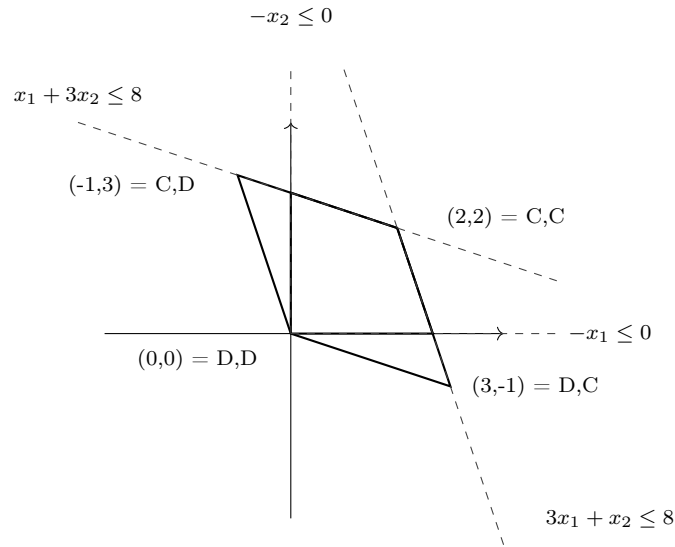
\begin{figure}[H]
\centering
\begin{tikzpicture}[scale=0.7]

  \draw[thick,black] (2.0,2.0) -- (-1,3) -- (0,0) -- (3,-1) -- (2,2);
  \draw[thick,black] (0,0) -- (2.67,0) -- (2,2) -- (0,2.67) -- (0,0);
  \draw[thin, ->] (-3.5,0) -- (4,0);
  \draw[thin, ->] (0,-3.5) -- (0,4);
  \draw[dashed, darkgray](-4,4)--(2.0, 2.0)--(5, 1);
  \draw[dashed, darkgray](4,-4)--(2.0, 2.0)--(1, 5);
  \draw[dashed, darkgray](0,0)--(0,5);
  \draw[dashed, darkgray](0,0)--(5,0); 

  \node[right,outer sep = 10.5pt, fill = white, font = \scriptsize] at (2.2,2.2)   (b) {(2,2) = C,C};
  \node[right,outer sep = 4pt, fill = white,font = \scriptsize] at (3,-1) (d) {(3,-1) = D,C};
  \node[left, outer sep = 12pt, fill = white,font = \scriptsize] at (-1,2.8) (d) {(-1,3) = C,D};
  \node[left, outer sep = 8.5pt, fill = white, font = \scriptsize] at (0,-0.5) (d) {(0,0) = D,D};

  \node at (-4,4.5) [font=\scriptsize] {$x_1 + 3x_2 \leq 8$};
  \node at (6,-3.5) [font=\scriptsize] {$3x_1 + x_2 \leq 8$};
  \node at (0,6) [font=\scriptsize] {$-x_2 \leq 0$};
  \node at (6,0) [font=\scriptsize] {$-x_1 \leq 0$};

\end{tikzpicture}
\caption{Linear inequalities that represent the individually rational feasible set}
\label{fig:h_representation}
\end{figure}

The main point here is to demonstrate the key iteration step from \hyperref[sec:key_iteration_step]{Section 4}. 
Consider the above introduced prisoners' dilemma and the signal structure, and assume the players are fairly patient with $\delta = \frac{9}{10}$.
We begin with the (individually rational) feasible payoff set $W^0$ in its V-representation. To accommodate the game's incentive constraints, we transform the set to its H-representation as shown in Figure \ref{fig:h_representation}.

\begin{figure}[htp]
\centering
\begin{subfigure}{0.5\textwidth}
  \centering
  \includegraphics[width=0.9\linewidth]{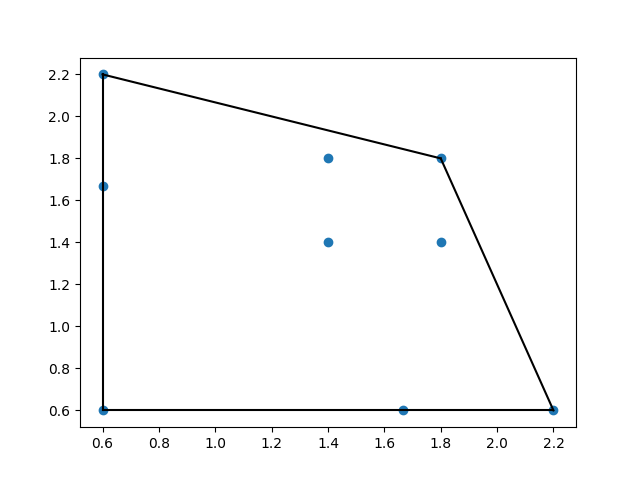}
  \caption{Play CC}
  \label{fig:sub1-cc}
\end{subfigure}%
\begin{subfigure}{0.5\textwidth}
  \centering
  \includegraphics[width=0.9\linewidth]{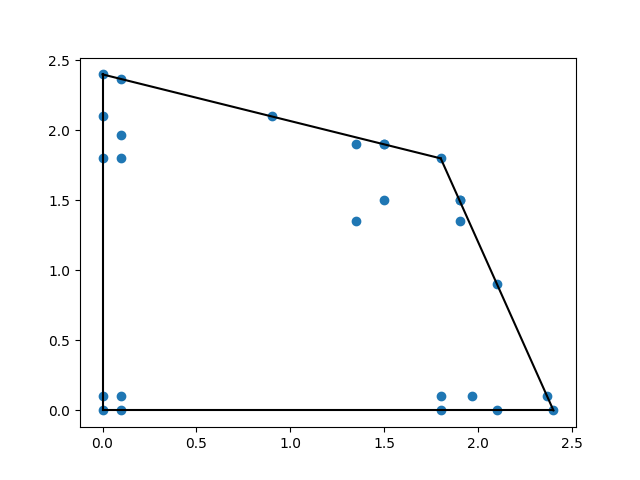}
  \caption{Play DD}
  \label{fig:sub2-dd}
\end{subfigure}

\begin{subfigure}{0.5\textwidth}
  \centering
  \includegraphics[width=0.9\linewidth]{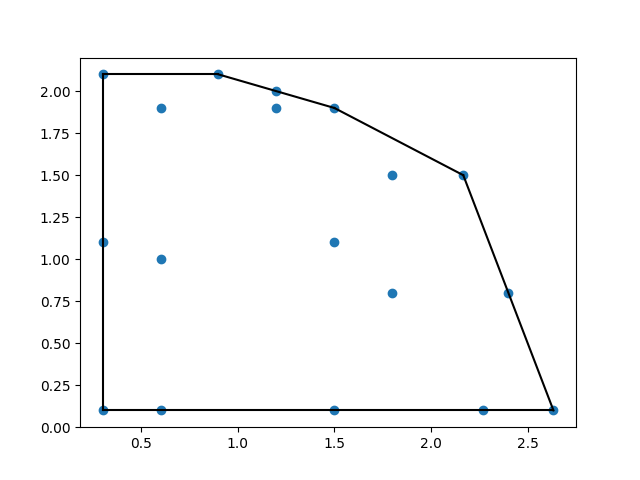}
  \caption{Play DC}
  \label{fig:sub3-dc}
\end{subfigure}%
\begin{subfigure}{0.5\textwidth}
  \centering
  \includegraphics[width=0.9\linewidth]{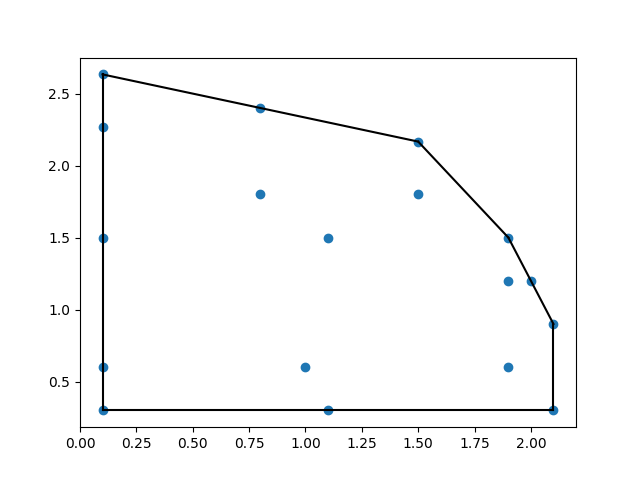}
  \caption{Play CD}
  \label{fig:sub4-cd}
\end{subfigure}
\caption{Possible repeated game payoffs after current play in iteration 1 for different current plays.}
\label{fig:continuation_payoffs}
\end{figure}

\begin{figure}[htp]
    \centering
    \includegraphics[width=9cm]{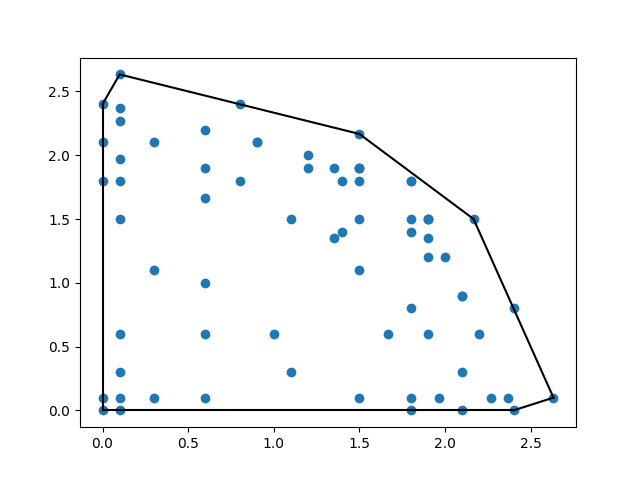}
    \caption{The set $W^1=\mathcal{B}(W^0)$ - the outcome payoff set of the first iteration.}
    \label{fig:first_iteration}
\end{figure}

The core of the iteration step is the decomposition of payoffs: $v_i$ is split into current payoffs $u(a)$ and continuation payoffs $\gamma (y)$.
These continuation payoffs are selected from $W^0$, which essentially outlines the range of future payoffs a player can expect by choosing a particular action today. 
Figure \ref{fig:continuation_payoffs} illustrates the collection of potential future payoffs that result from current action choices and considering the incentive constraints, using the algorithm. The first iteration produces a new set, $\mathcal{B}(W^0)$, presented in its V-representation (Figure \ref{fig:first_iteration}), which then becomes the input for the next iteration. 

\clearpage 

The next (second) iteration then takes the set of extreme points of the set $W^1=\mathcal{B}(W^0)$ as its input and produces (the set of extreme points of) $W^2=\mathcal{B}(W^1)$, with generally the $k$-th iteration producing (the set of extreme points of) $W^k=\mathcal{B}(W^{k-1}) = \mathcal{B}^{k-1}(W^0)$. 

Having detailed the key iteration step, an error margin, here $\epsilon = 0.005$, is set to quantify the differences in the areas between two successive $W^{k-1}$ and $W^k$ such that if this area difference falls below $\epsilon$ the algorithm stops. In this case, we also say that the algorithm converged. 

\begin{figure}
    \centering
    \includegraphics[width=0.7\textwidth]{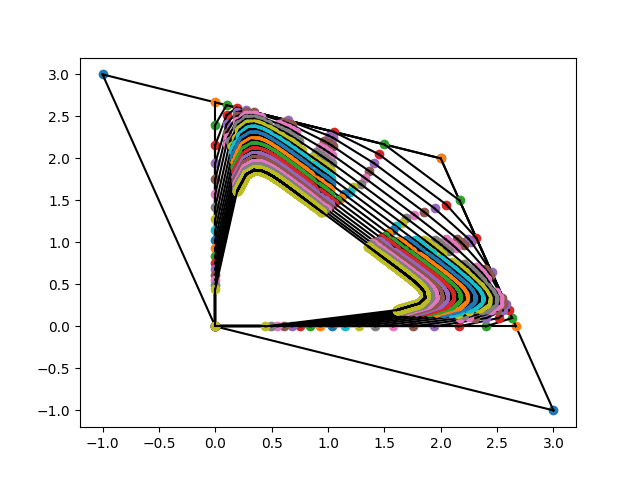}
    \caption{The payoff set after 17 iterations, $\delta = \frac{9}{10}$, without the RDP feature.}
    \label{fig:17_rounds}
\end{figure}

\begin{figure}
    \centering
    \includegraphics[width=0.7\textwidth]{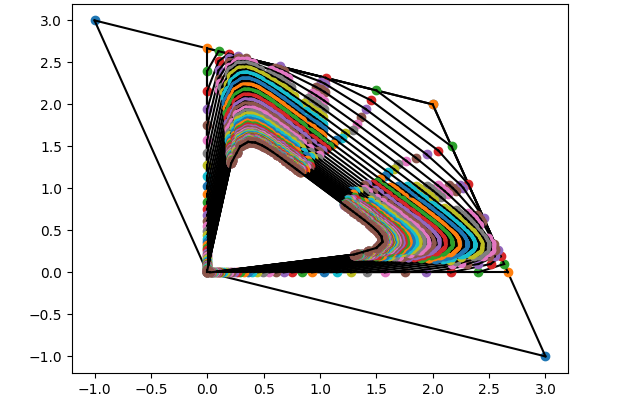}
    \caption{The payoff set after 44 iterations, $\delta = \frac{9}{10}$, with the RDP feature.}
    \label{fig:44_iteration}
\end{figure}

The execution of the algorithm shows an expected exponential growth in extreme points, as noted by \cite{abreu1990toward}. The growth poses computational challenges, especially noticeable in later iterations. Enabling the RDP feature results in a notable improvement in computation time, allowing us to focus on results rather than the computation process.\footnote{Without the RDP optimization, initial algorithms' computation time is approximately 64 hours to reach the 17th iteration, illustrating the intensive computational demand (see Figure \ref{fig:17_rounds}). With the RDP feature enabled, this computation time was drastically reduced to approximately 10 minutes. All parameters are kept the same.} With the RDP feature, the algorithm reaches convergence by the 44th iteration (Figure \ref{fig:44_iteration}).
 
\begin{figure}
    \centering
    \includegraphics[width=0.7\textwidth]{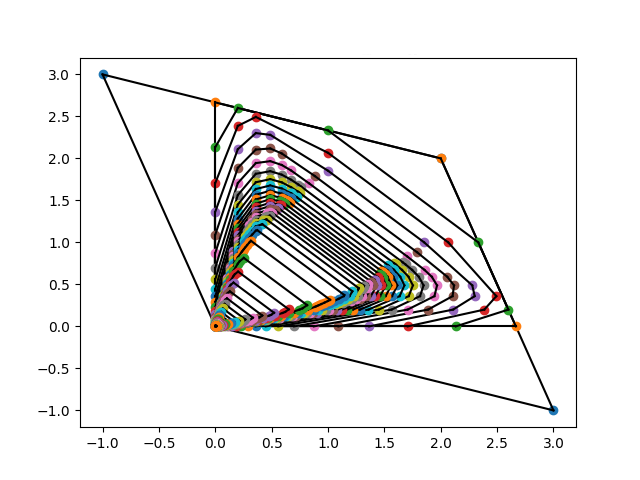}
    \caption{The payoff set converges to the stage game Nash equilibrium if $\delta = \frac{8}{10}$ (or lower).}
    \label{fig:nash_equilibrium}
\end{figure}

Figure \ref{fig:nash_equilibrium} considers the discount factor $\delta = \frac{8}{10}$ and the stage game Nash equilibrium is the only sustainable outcome. This is also the result for any $\delta \leq \frac{8}{10}$.

\subsection{A simplified Cournot game}

Real-world strategic interactions often extend beyond the simple prisoners' dilemma framework and two observable signals. To better capture the complexity of these interactions, we explore models with richer action and signal spaces. 
Drawing from the seminal works of \cite{judd2003computing} and \cite{abreu2014algorithm}, consider a Cournot duopoly stage game with a comprehensive payoff matrix given in Table \ref{fig:cournot_payoff_matrix}.

\begin{table}[htb]
    \centering
    \begin{center}
    \begin{tabular}{c|ccc}
        & L & M & H \\
        \hline
        L & $16, 9$ & $3, 13$ & $0,3$ \\
        
        M & $21,1$ & $10,4$ & $-1,0$ \\
        
        H & $9,0$ & $5,-4$ & $-5,-15$ \\
        
    \end{tabular}
    \end{center}
    \caption{Cournot duopoly payoff matrix}
    \label{fig:cournot_payoff_matrix}
\end{table}

Accompanying the payoff matrix is a comprehensive signal structure that categorizes outcomes based on four publicly observable signals, as given in Table \ref{table:signal_structure}. 

\begin{table}[htb]
\centering
\renewcommand{\arraystretch}{0.99} 
\begin{tabular}{l|p{3cm}p{3cm}p{3cm}p{3cm}}
    
    & $y_1$ & $y_2$ & $y_3$ & $y_4$ \\
   \hline
    L,L & 1 & 0 & 0 & 0 \\
    
    L,M &  0.5 & 0.5 & 0 & 0 \\
    
    L,H & 0 & 1 & 0 & 0 \\
    
    M,L & 0.5 & 0 & 0.5 & 0 \\
    
    M,M & 0.25 & 0.25 & 0.25 & 0.25 \\
    
    M,H & 0 & 0.5 & 0 & 0.5 \\
    
    H,L & 0 & 0 & 1 & 0 \\
   
    H,M & 0 & 0 & 0.5 & 0.5 \\
    
    H,H & 0 & 0 & 0 & 1 \\
    
\end{tabular}
\caption{The signal structure}
\label{table:signal_structure}
\end{table}

\begin{figure}
    \centering
    \includegraphics[width=0.6\textwidth]{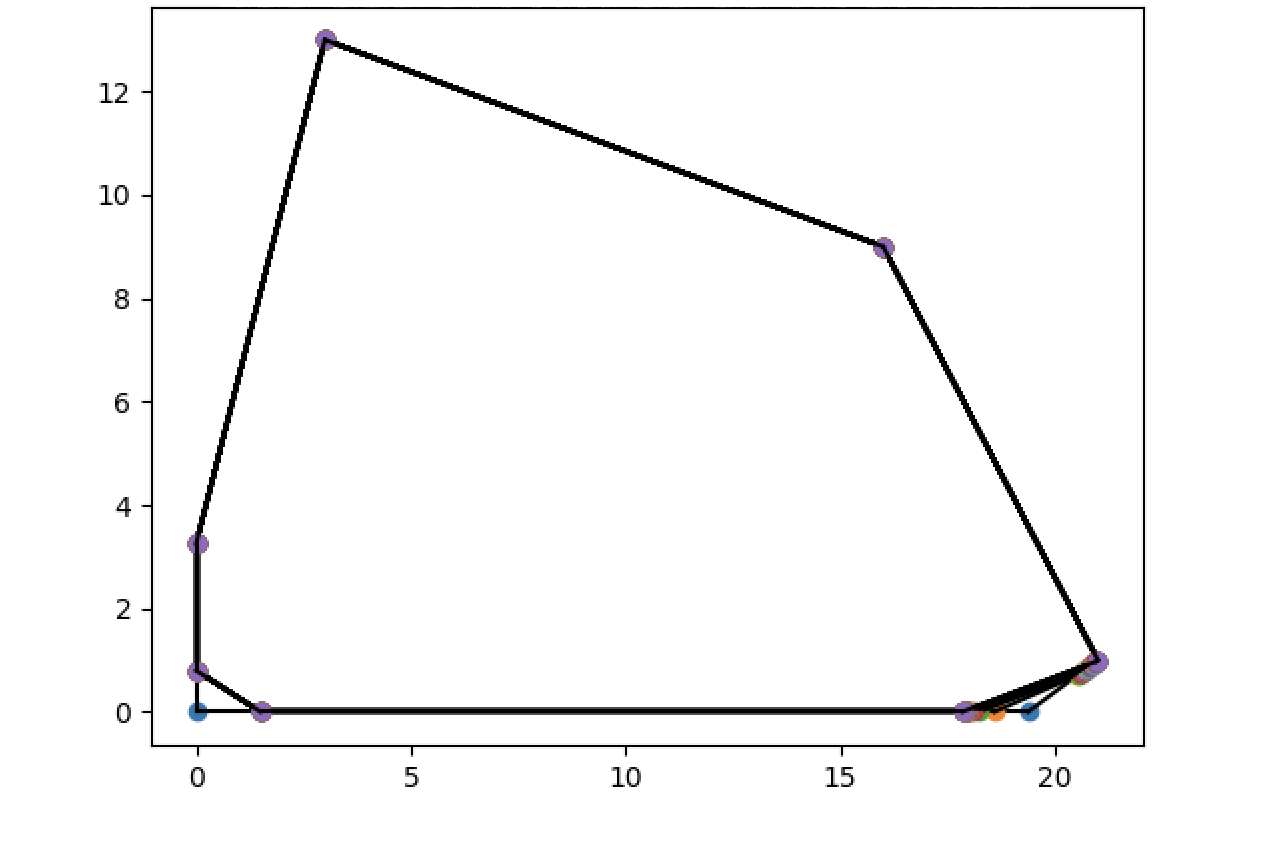}
    \caption{Three-by-three Cournot duopoly for $\delta = \frac{9}{10}$.}
    \label{fig:cournot_patient}
\end{figure}

\begin{figure}
    \centering
    \includegraphics[width=0.6\textwidth]{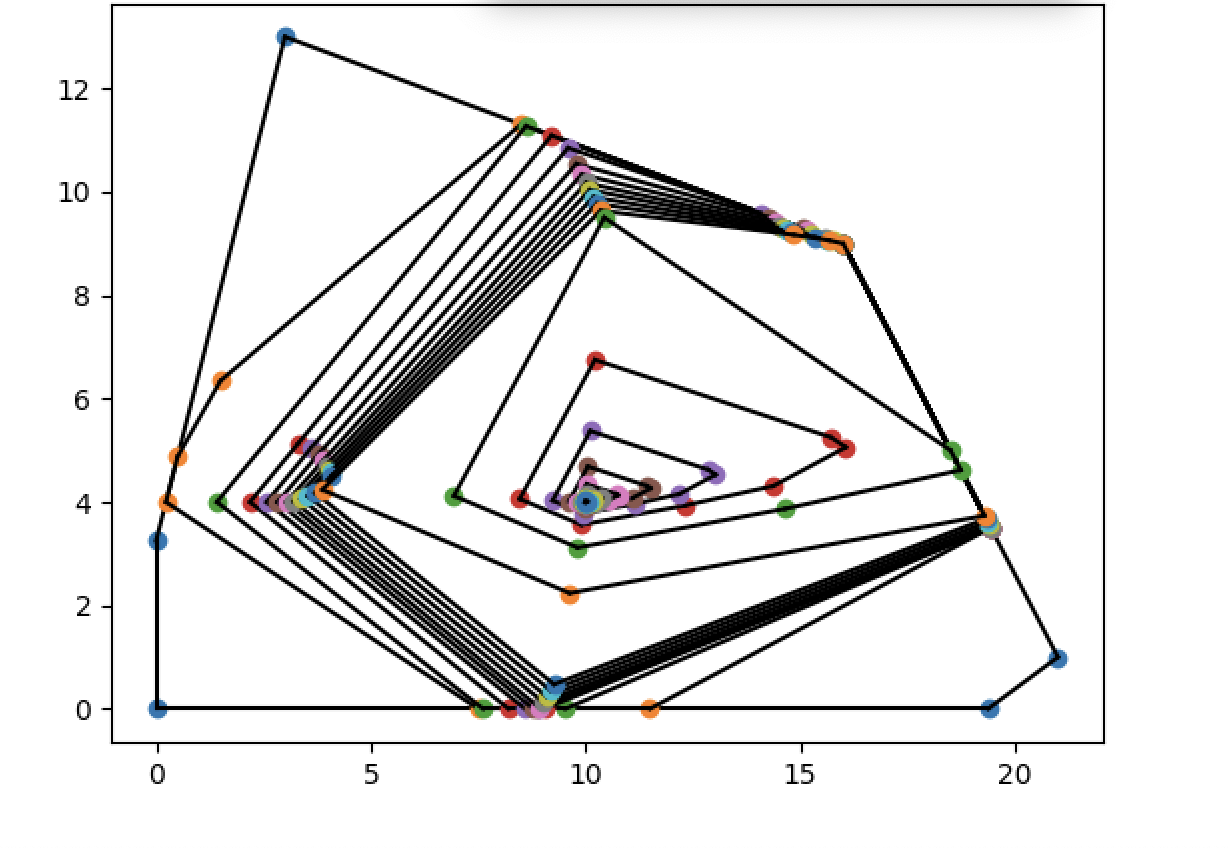}
    \caption{The payoff set converges to the stage game Nash equilibrium if $\delta = \frac{1}{2}$ (or lower).}
    \label{fig:cournot_impatient}
\end{figure}

The analysis of the Cournot example is similar to that of the previous example. Assume again that players are patient, i.e., $\delta = \frac{9}{10}$, the signal structure is the one from Table \ref{table:signal_structure} and the error margin is $\epsilon = 0.005$. Figure
\ref{fig:cournot_patient} illustrates the equilibrium payoff set that converged at iteration 23.
The payoff set closely approximates the entire individually rational feasible set. This suggests that patient players achieve more efficient outcomes, even under imperfect monitoring conditions. Conversely, when the discount factor falls below the $\delta = \frac{1}{2}$, the sustainable outcomes are restricted to the stage game Nash equilibrium (see Figure \ref{fig:cournot_impatient}). 

\newpage
The examples in this section demonstrate that determining equilibrium payoff sets in repeated games under imperfect public monitoring is not straightforward; the patience parameter matters and the algorithm successfully captures variations in different $\delta$ values.

\newpage
\section{Discussion}
\label{sec:discussion}
This paper introduces an algorithm to calculate an upper bound of the set of perfect public equilibrium payoffs for two-player repeated games with imperfect public monitoring, public randomization, and discounting. Building on the framework established by \citet*{abreu1990toward}, it offers an explicit method for computing payoff sets for any given discount factor, $\delta$.

The core challenge addressed by our algorithm is defining an iteration step that refines a convex payoff set, and repeats this iteration step until convergence is achieved. 
Initially, the algorithm starts with the V-representation (the set of extreme points of a convex set), capturing the feasible and individually rational payoffs, and then transforms the set to its H-representation (representing a convex set by its collection of bounding hyperplanes), to account for the incentive constraints. 

While \cite{sugaya2023monitoring}, \cite{abreu1991information}, \cite{sannikov2010role} establish conditions under which equilibrium payoffs can be supported with impatient players, this algorithm provides an upper bound for equilibrium payoff sets with fixed monitoring structure for any value of the discount factor $\delta<1$.

The current algorithm is fully functional for two-player repeated games, with each player having up to three actions, and computationally efficient for games with up to four signals.
Future research could provide an inner approximation of the payoff set and extend the algorithm to games with more than two players. 
Additionally, exploring more efficient computational techniques to manage the exponential growth of extreme points is another potential direction.

\newpage

\bibliography{bibliography.bib}

\end{document}